\documentclass[aps,pre,
              amsmath,amssymb,amsfonts,groupedaddress,
              twocolumn,
              showpacs,
              superscriptaddress,floatfix]{revtex4-1}

\usepackage{graphicx}
\usepackage{dcolumn}
\usepackage{bm}

\usepackage{color}

\newcommand{\Wcm}{\,{\rm W \, cm^{-2}}}
\newcommand{\beq}{\begin{equation}}
\newcommand{\eeq}{\end{equation}}
\newcommand{\beqa}{\begin{eqnarray}}
\newcommand{\eeqa}{\end{eqnarray}}
\newcommand{\bma}{\mathbf}
\newcommand{\mum}{\,{\mu \rm m}}
\newcommand{\MeV}{\,{\rm MeV}}
\newcommand{\fes}{\,{\rm fs}}

\begin{document}

\title{Radiation pressure dominant acceleration: polarization and\\ radiation reaction effects, and energy increase in three-dimensional simulations}

\author{M. Tamburini}\email{tamburini@df.unipi.it}
\affiliation{Istituto Nazionale di Ottica, CNR, research unit ``A. Gozzini'', Pisa, Italy}
\affiliation{Dipartimento di Fisica ``E. Fermi'', Universit\`a di Pisa, Largo Bruno Pontecorvo 3, I-56127 Pisa, Italy}
\author{T. V. Liseykina}
\affiliation{Institut f\"ur Physik, Universit\"at Rostock, D-18051 Rostock, Germany}
\author{F. Pegoraro}
\affiliation{Dipartimento di Fisica ``E. Fermi'', Universit\`a di Pisa, Largo Bruno Pontecorvo 3, I-56127 Pisa, Italy}
\affiliation{Istituto Nazionale di Ottica, CNR, research unit ``A. Gozzini'', Pisa, Italy}
\author{A. Macchi}
\affiliation{Istituto Nazionale di Ottica, CNR, research unit ``A. Gozzini'', Pisa, Italy}
\affiliation{Dipartimento di Fisica ``E. Fermi'', Universit\`a di Pisa, Largo Bruno Pontecorvo 3, I-56127 Pisa, Italy}

\date{\today}

\begin{abstract}
Polarization and radiation reaction (RR) effects in the interaction of a 
superintense laser pulse
($I > 10^{23} \Wcm$) with a thin plasma foil are investigated with three 
dimensional particle-in-cell (PIC) simulations.
For a linearly polarized laser pulse, strong anisotropies such as the 
formation of two high-energy clumps
in the plane perpendicular to the propagation direction and significant 
radiation reactions effects are observed.
On the contrary, neither anisotropies nor significant radiation reaction 
effects are observed using circularly polarized laser pulses,
for which the maximum ion energy exceeds the value obtained in 
simulations of lower dimensionality.
The dynamical bending of the initially flat plasma foil leads to
the self-formation of a quasi-parabolic shell that focuses the impinging 
laser pulse strongly increasing its energy and momentum densities.
\end{abstract}

\pacs{52.38.Kd, 52.65.Rr}

\maketitle

The radiation pressure generated by ultraintense laser pulses may drive
strong acceleration of dense matter, as experimentally 
shown in various regimes \cite{karPRL08b,*akliPRL08,*henigPRL09b,*palmerPRL11}.
Radiation pressure may thus be an effective mechanism for the generation of 
high-energy ions, especially in the regime of extremely high 
intensities and relativistic ion energies as foreseen with the ELI
project. 
In the case of solid-density thin foil targets, pioneering 
particle-in-cell (PIC) simulations showed that at 
intensities exceeding $10^{23} \Wcm$ and for linear polarization of the laser pulse
radiation pressure dominates the acceleration 
yielding linear scaling with the laser pulse intensity, high efficiency and
quasi-monoenergetic features in the ion energy spectrum 
\cite{esirkepovPRL04}. 
More recent two-dimensional (2D) simulations for 
a small disk target suggested a potentially ``unlimited'' energy gain for the fraction
of ions that get phase-locked with the laser pulse \cite{bulanovPRL10}.

The above mentioned studies showed that the radiation pressure 
dominant acceleration (RPDA) regime is very appealing as a route to the 
generation of relativistic ions, but leave several theoretical issues open.
Firstly, transverse instabilities \cite{pegoraroPRL07} and 
multi-dimensional effects may play a crucial role
as shown by 2D simulations \cite{bulanovPRL10}.
Secondly, the use of circular polarization (CP) instead of linear polarization (LP)
quenches the generation of high-energy electrons \cite{macchiPRL05},
allowing radiation pressure to dominate even at intensities below $10^{23} \Wcm$
and leading to efficient acceleration of ultrathin foils \cite{zhangPP07,*robinsonNJP08,*klimoPRSTAB08};
it has not been shown yet whether the use of CP is advantageous also
at ultra-high intensities ($I > 10^{23} \Wcm$), i.e. when the radiation pressure
generated by the laser pulse becomes the dominant mechanism of acceleration both for CP and LP.
Finally, it has been shown by one-dimensional (1D) simulations that radiation reaction (RR) effects
may significantly affect the dynamics of radiation pressure acceleration
both for thick \cite{naumovaPRL09} and thin targets \cite{tamburiniNJP10,chenPPCF11},
and also depend strongly on the polarization \cite{tamburiniNJP10}.
All of these phenomena may be affected by the dimensionality of the problem,
and a fully three-dimensional (3D) approach is ultimately needed because e.g. 
in 2D simulations and for LP the laser-plasma coupling is 
different for $S$- and $P$-polarization (i.e. for the electric field
of the laser pulse either perpendicular or parallel to the simulation plane, respectively)
and the constraint of the conservation of angular momentum carried by 
CP pulses holds in 3D only.

In this paper, we address the role of polarization and RR effects
in the RPDA regime using fully 3D PIC simulations.
To our knowledge, these are the first 3D simulations
of ion acceleration in the RPDA regime with RR effects included
and among the largest and most accurate 3D simulations reported so far. 
Our results show that even in the RPDA regime CP leads to higher ion energies
and better collimation than LP, for which an anisotropic ion
distribution is observed. 
It is also found that the bending of the foil leads to
a self-generated parabolic shell that focuses the impinging pulse down to
an almost $\lambda^3$ volume and that the energy density at the focus
largely exceeds the initial peak energy density. 
Compared to 2D simulations with analogous parameters,
the pulse focusing effect is remarkably enhanced and the cut-off energy of ions 
is increased. RR effects on the ion spectrum are found to be negligible for
CP but quite relevant for LP where they increase the energy cut-off.

\begin{figure*}[th!]
\begin{center}
\includegraphics[width=0.98\textwidth]{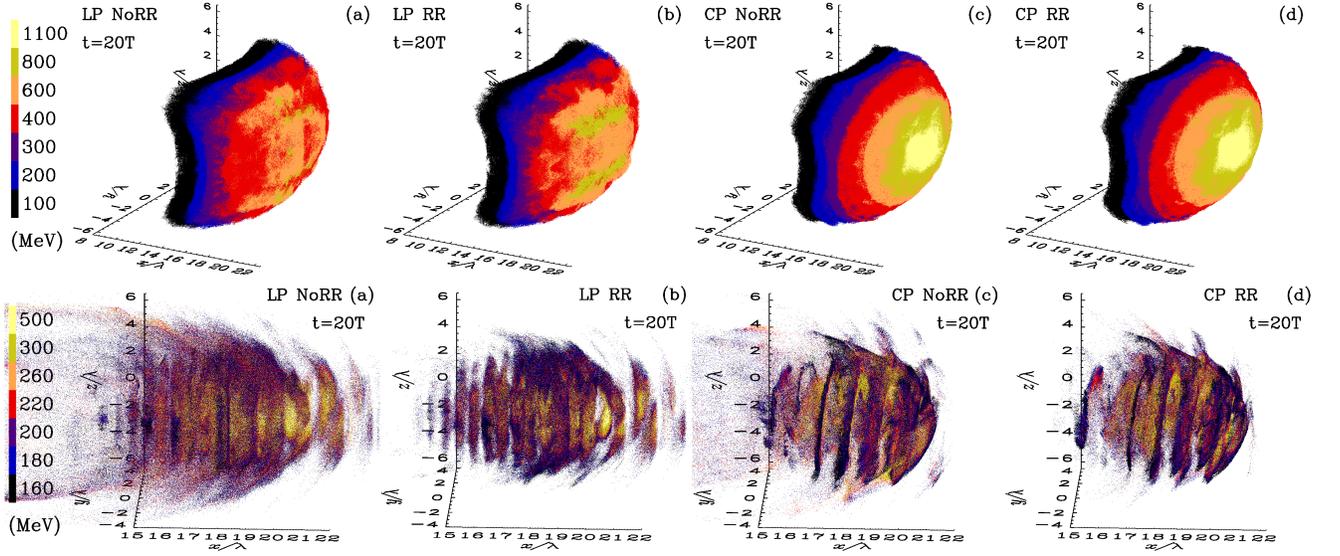}
\caption{Spatial  distributions of ions (upper row) and electrons (lower row) 
at $t = 20 \, T$ and in the region $(|y|,|z|) \leq 5.7\lambda$, 
for LP without (a) and with (b) RR
and for CP without (c) and with (d) RR.
Ions and electrons are divided into seven populations according to their 
kinetic energy, with the color-bar reporting the lower bound of the  
energy interval. 
In the LP case (frames (a,b)), the polarization is along the $y$ axis.} 
\label{3D}
\end{center}
\end{figure*}

Our approach is based on the numerical solution of kinetic equations for
the phase space distributions of electrons and ions, where RR is included
in the motion of electrons via the Landau-Lifshitz (LL) force 
\cite{landau-lifshitz2}. Details of our RR modeling 
and numerical implementation in a PIC code
are given in Refs.\cite{tamburiniNJP10,tamburiniNIMA11}.
The effective equation of motion for electrons, 
after neglecting terms which are negligible 
in the classical limit \cite{tamburiniNJP10} is
\begin{eqnarray}
\frac{d{\bf p}}{dt}&=&-e\left({\bf E} + \frac{\bf v}{c} \times {\bf B} \right)
                    +{\bf f}_R , \nonumber \\
{\bf f}_R&=& \frac{2r_c^2}{3}
\left\{-\gamma^2
\left[ \left( {\bf E} + \frac{\bf v}{c} \times {\bf B} \right)^2
- \left( \frac{\bf v}{c} \cdot {\bf E} \right)^2  \right] \frac{\bf v}{c}   
\right.
\nonumber \\
&  & \left. +
\left[ \left( {\bf E} + \frac{\bf v}{c} \times {\bf B} \right) \times {\bf B} +
\left( \frac{\bf v}{c} \cdot {\bf E} \right) {\bf E} \right]
\right\} , 
\label{3dLL}
\end{eqnarray}

where $r_c=e^2/m_e c^2$.
The RR force contribution is important for ultrarelativistic electrons and is
usually dominated by the first term, while the second term
ensures the on-shell condition \cite{tamburiniNJP10}. 
Notice that the dominant term has almost the same form
also in different approaches to RR modeling \cite{naumovaPRL09,chenPPCF11}.
For a plane wave propagating along the $x$-axis, the RR force is maximum
or zero for counterpropagating ($v_x/c \rightarrow -1$) or copropagating
($v_x/c \rightarrow +1$) electrons, respectively.

In order to clarify the new qualitative features due to RR effects,
we recall that the phase space volume element $J$ evolves according to
$dJ/dt= J\nabla_{\bma{p}}\cdot\bma{f}_R$.
It has been shown \cite{tamburiniNIMA11} that 
$\nabla_{\bma{p}}\cdot\bma{f}_R\leq 0$
and therefore the RR force leads to a \emph{contraction} of the available 
phase space volume.
The physical interpretation of this property is that the RR force
acts as a cooling mechanism for the system accounting for
the emission of high-energy photons. These photons are assumed to escape
from the plasma freely, carrying away energy 
and entropy \cite{tamburiniNIMA11}.

We present a total of four 3D simulations each with
the same physical and numerical parameters but
different polarization, with and without RR effects. 
In these simulations, the laser field amplitude has
a $\sin^2$-function longitudinal profile with $9\lambda$ FWHM
(where $\lambda=0.8~\mum$ is the laser wavelength) 
while the transverse radial profile is Gaussian with $10\lambda$ FWHM
and the laser pulse front reaches the edge of the plasma foil at $t=0$.
The peak intensity at the focus is $I = 1.7 \times 10^{23} \Wcm$ which 
corresponds to a normalized amplitude $a_0=280$ for LP and $a_0=198$ for CP.
The target is a plasma foil of electrons and protons
with uniform initial density $n_0 = 64n_c$ (where
$n_c = \pi m_ec^2/e^2\lambda^2$ is the critical density),
thickness $\ell = 1 \lambda$ and initially located
in the region $10\lambda \leq x \leq 11 \lambda$.
The density $n_0 \simeq 1.1 \times 10^{23}~\mbox{cm}^{-3}$
is slightly lower than that of solid targets
but the areal density $n_0\ell$ has fully realistic values. 
Moreover, laser pulses of ultrahigh contrast are now available
\cite[and references therein]{thauryNP07,*rodelAPB11}
to avoid early plasma formation effects by the prepulse, 
thus a thin plasma with step-like profile is not an unrealistic assumption.

The simulation grid is $1320\times 896\times 896$
and the spatial step is $\lambda/44$ for each direction. 
The timestep is $T/100$ where $T=\lambda/c=2.67 \fes$ is the 
laser period.
We use $216$ particles per cell for each species and the total number of 
particles is $1.526 \times 10^{10}$.
The runs were performed using 1024 processors each one equipped
with 1.7 GB of memory of the IBM-SP6 cluster at the CINECA
supercomputing facility in Bologna, Italy.

\begin{figure*}[t!]
\begin{center}
\includegraphics[width=0.98\textwidth]{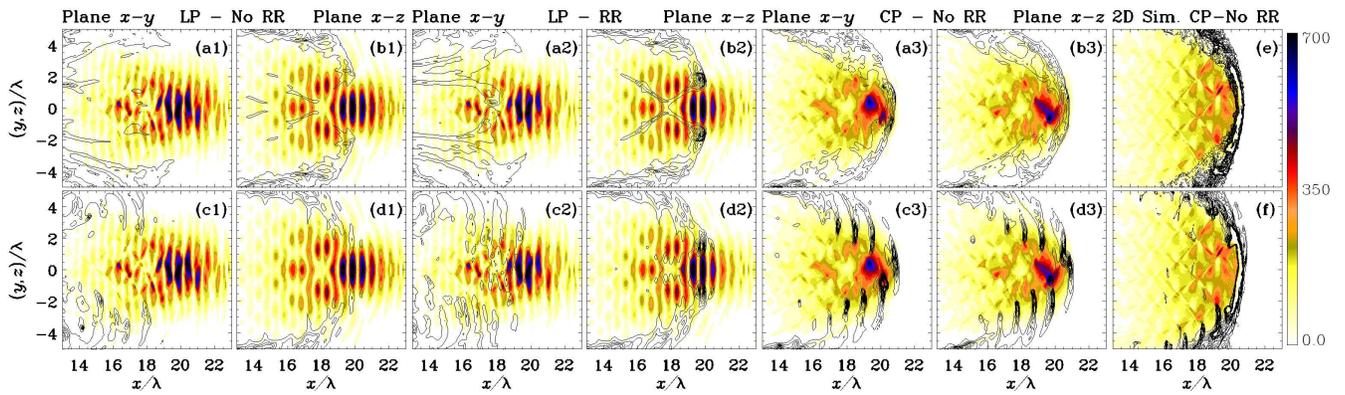}
\caption{($x,y$) and ($x,z$) sections of the 3D simulations
of the laser pulse-foil interaction [a1-3), b1-3), c1-3), d1-3)]
and 2D simulations for CP and without RR with the same parameters
as the 3D simulations [e), f)], all at $t = 20 \, T$.
Each frame reports the color contours of $\sqrt{\bma{E}^2+\bma{B}^2}$
(normalized units) in the $x-y$ plane at $z=0$ (a1-3 and c1-3), 
in the $x-z$ plane
at $y=0$ (b1-3 and d1-3) and in the simulation plane e), f) in the 2D case.
Line contours of the ion and electron densities are superimposed
in the upper and lower frames, respectively.
The CP case with RR is almost identical to the CP case without RR and
it is not reported.} \label{2DSections}
\end{center}
\end{figure*}

Fig.\ref{3D} shows the ion and the electron 3D spatial distributions at 
$t=20 \, T$ for the LP case without (a) and with (b) RR and for the CP case 
without (c) and with (d) RR. The color corresponds to the range in kinetic 
energy. 
For CP, the ion spatial distribution
follows the spatial intensity profile of the initial laser pulse,
has rotational symmetry around the central axis, 
and a distribution in energy monotonically decreasing with increasing 
radial distance.
The most energetic ions are located near the axis.
The number of ions having energy ${\cal E} \geq 1100 \MeV$ 
and ${\cal E} \geq 800 \MeV$ 
are $2.3 \times 10^{10}$ and $9.4 \times 10^{10}$, respectively.
The electron spatial distribution has an helicoidal shape
with step $\lambda$ Fig.\ref{3D}~c),d). 
RR effects play a minor role for CP affecting only a small fraction
of ultra-relativistic electrons 
(mostly removing fast electrons \emph{behind} the foil
with almost no influence on the ion distribution
as seen by comparing frames~c) and d) in Fig.\ref{3D}). 

The ($x,y$) and ($x,z$) sections of the total electromagnetic energy density and
of the ion and electron densities for the CP case in 
Fig.\ref{2DSections}~a3)-b3) and c3)-d3) evince a 
self-generated parabolic shell wrapping the laser pulse 
and focusing it up to nearly a $\lambda^3$ volume, so that 
both the energy and the momentum densities
at the focus reach values more than eight times their peak value in the initial 
laser pulse. This effect is much 
weaker in 2D simulations with the same parameters
as shown in Figs.\ref{2DSections}~e)-f).
Along the axis, the peak value and width of the ion density profile
are $\simeq 10n_c$ and $\simeq 0.5\lambda$, showing a strong rarefaction
due to the transverse expansion, potentially leading to enhanced acceleration
as described in Ref.\cite{bulanovPRL10}.

For LP, the peak ion energy is lower than for CP, the ion distribution
is anisotropic and RR effects are much stronger.
The most energetic ions (800-1100 \MeV)
are grouped into two off-axis clumps lengthened and aligned
along the polarization direction, and their number is increased
in the case with RR as seen by the comparison of Figs.\ref{3D}~a) and b) and also 
in Figs.\ref{2DSections}~a1),a2) and b1),b2) 
where sections of the ion density
in the ($x,y$) and ($x,z$) planes are shown.
The contours of the electromagnetic (EM) energy density in 
Figs.\ref{2DSections}~a1),a2) and b1),b2) show that near the axis most of the 
laser pulse has been transmitted 
through the target. The increased bunching and higher density observed in 
the case with RR may be related to the higher ion energies since the local 
increase of the density and therefore of the reflectivity leads to 
a longer and more efficient RPDA phase. This is consistent with observing 
in Figs.\ref{2DSections}~b1) and b2) that the EM energy density is higher 
behind the two high-density clumps, which correspond to the most energetic 
ions and are similar to
the ion lobes observed in Ref.\cite{yinPRL11} at lower intensity and in 
a regime of strong pulse penetration through the foil.
The pulse focusing effect by the self-generated parabola is present 
also in the LP case although weaker than in the CP case, and presumably
reduced as the laser pulse breaks through the parabolic shell.

The differences between CP and LP in the acceleration dynamics
are well explained, for planar geometry and non-relativistic ions,
by the absence of the oscillating component of the
$\bma{J} \times \bma{B}$ force for CP \cite{macchiPRL05}, which
maximizes the effect of the radiation pressure strongly suppressing the electron heating. 
The absence of the oscillating component of the
$\bma{J} \times \bma{B}$ force also accounts for the very
different RR effects. 
For CP, a steady push of the foil and weak pulse penetration are observed,
most of the electrons move coherently with the foil
and in the same direction as the laser pulse so that
the RR force becomes very small  
in accordance with Eq.(\ref{3dLL}) since the electrons effectively
copropagate with the laser pulse (see also
\cite{esirkepovPRL04,tamburiniNJP10}).
For LP, the $\bma{J} \times \bma{B}$-driven oscillations
allow electrons to collide with the counterpropagating laser pulse
twice per cycle producing temporal maxima in the RR force 
in agreement with Eq.(\ref{3dLL}).
Our present results indicate that CP leads to more efficient 
acceleration, producing higher energy and collimated ion beams, and 
making RR effects negligible also in the 3D case, accounting for target
bending and pulse focusing effects, and for relativistic ions.

\begin{figure}[b]
\begin{center}
\includegraphics[width=0.48\textwidth]{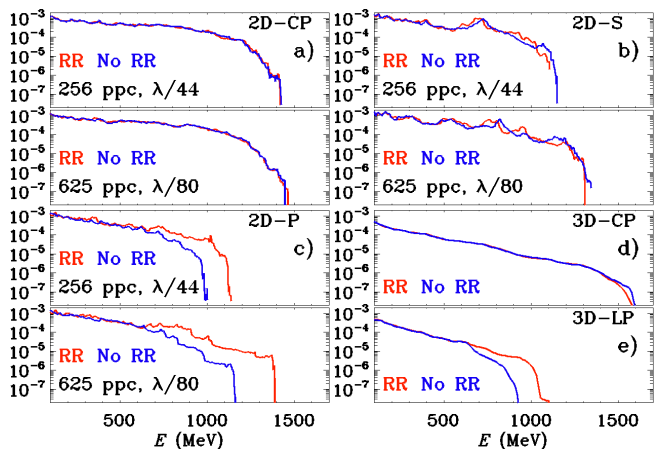}
\caption{Ion spectra from 2D [a)-c)] and 3D [d)-e)] simulations with
same physical parameters, all at $t=20T$.
The 2D spectra are reported for circular (CP, frame a)) and 
linear (LP) ``$S$'' (frame b)) and ``$P$'' (frame c)) polarization cases.
In each plot, the blue and red curves correspond to simulations
without and with radiation reaction (RR) effects, respectively.
In the upper plots of frames a)-c) the numerical resolution (number of 
particles per cell and of points per wavelength) 
is similar to those of the 3D simulations in d)-e), while in the 
lower plots the results for higher resolution are shown.}
\label{spectra}
\end{center}
\end{figure}

Large-scale 3D PIC simulations are limited
by the size and availability of computational resources both 
in the number of runs that may be performed and
in the achievable numerical resolution.
This last issue may raise doubts on the accuracy of 3D results.
To gain confidence on this side, as well as to compare the 3D results
to those obtained in lower dimensionality,
we performed 2D simulations both with numerical parameters similar
to those of 3D runs and with higher resolution.
The effect of increasing resolution and particle number on the ion spectra 
in 2D simulations is shown in Fig.\ref{spectra} where 2D results are 
reported for the three different polarization cases 
(CP, LP-$S$ and LP-$P$) and compared with the 3D results for both LP and CP.
The spectra are normalized to unity both in the 2D and 3D cases.
In the CP case both numerical and RR effects
on the spectrum are smaller while in the $P$-polarization case
these effects are larger.
Changing the spatial resolution from $\lambda/44$
to $\lambda/80$ and increasing the number of particles-per-cell (ppc)
for each species from $256$ to $625$ shifts
the energy cut-off by $\sim 2\%$ in the CP case
and by $\sim 15\%$ $(\sim 20\%)$ in the $P$-polarization case without RR 
(with RR). 
The stronger effect of the inclusion of RR for the higher resolution case
may be explained by noticing that RR mostly affects the highest energy 
electrons \cite{tamburiniNJP10}, which are located
in the high-energy tail of the distribution function that needs a very large 
number
of particles to be resolved properly.
Nevertheless, the limited resolution does not qualitatively affect 
prominent features in ion spectra, such as the higher ion energy for CP and 
the relevance of RR effects for LP only, leading for this latter case to an 
\emph{higher} energy of ions with respect to the case of no RR
as observed in 1D simulations \cite{tamburiniNJP10}. 
As a novel feature of 2D simulations,
$P$-polarization leads to
much stronger RR effects than $S$-polarization. 
In fact, for $P$-polarization the electric field 
can drag a large fraction of electrons out in vacuum and towards the 
laser pulse as the plasma foil begins to bend, enhancing the RR effect. 

For CP, the maximum ion energy is higher in the 3D
case ($\simeq 1600~\MeV$) than in the 2D case ($\simeq 1400~\MeV$).
In turn, the latter value is higher than what found in 1D, plane-wave 
simulations, for which we find a broad spectral peak which at $t=20$ extends up to 
$\simeq 1100~\MeV$ and is centered around a value 
of $\simeq 870~\mbox{MeV}$. The latter value corresponds to the energy 
${\cal E}_{LS}=(\gamma -1) m_p c^2$, where $\gamma = 1/\sqrt{1-\beta^2}$
and $\beta$ are obtained from the ``light sail'' model \cite[and references therein]{macchiNJP10}
by numerically integrating the 1D equations of motion for the foil
\begin{eqnarray}
\frac{d\gamma \beta}{d t} = \frac{2 I (t - X/c)}{n_0 \ell m_p c^2} \left( \frac{1-\beta}{1+\beta} \right), 
\qquad 
\frac{d{X}}{dt}={\beta c} .
\end{eqnarray}
From the 1D modeling we also evaluate a final ion energy
for the spectral peak of $\simeq 1700~\MeV$ that is reached at $t \simeq 90$.
The 3D simulations could not be extended up to the end of the acceleration
stage (estimated up to $\sim 90~T$ in 1D simulations) but,
since the efficiency of RPDA increases with the foil velocity, the energy gain
is expected to be even larger at the later times (provided that 3D effects do not cause an 
early stop of the acceleration). 
Hence, the comparison at $t=20$ shows an overall \emph{increase}
of the ion energy in the 3D case with respect to 1D and 2D cases.
Part of the energy enhancement can be attributed to the reduction of
the areal density $n_0 \ell$ due to the transverse expansion, as was noticed in 2D 
simulations supporting the model of ``unlimited'' acceleration
\cite{bulanovPRL10}. An additional contribution may come from the above
described focusing of the laser pulse by the deformed foil, which is 
stronger in 3D geometry. This latter effect was absent in the 
simulations of Ref.\cite{bulanovPRL10} because a target with a radius 
smaller than the pulse waist was considered.

In conclusion, 
with three-dimensional particle-in-cell simulations 
of ultra intense laser interaction with solid-density foils
we showed that circular polarization improves ion acceleration also
in the radiation pressure dominant regime,
confirming and extending previous results obtained for lower intensity and/or
lower dimensionality.
In detail, circular polarization leads to the highest ion energies, to 
a symmetrical and collimated distribution, and to negligible effects
of radiation reaction. In addition,
the maximum energy of ions in 3D is larger than observed in corresponding
1D and 2D simulations.
This enhancement is attributed both to the density decrease in the
target, as noticed in the ``unlimited acceleration'' model 
\cite{bulanovPRL10}, and to the strong focusing of the laser pulse by the
parabolically deformed foil.
In the linear polarization case, lower maximum energies are achieved,
the most energetic ions are grouped anisotropically
into two off-axis clumps and
radiation reaction effects significantly affect the energy spectrum.
We expect these findings to be of relevance for the design of future 
experiments on laser acceleration of ions up to relativistic (GeV) energy.

\begin{acknowledgments}
Work sponsored by the Italian Ministry of University and  Research
via the FIRB project ``SULDIS''.
We acknowledge the CINECA Award N.HP10A25JKT-2010 for the availability of high 
performance computing resources.
\end{acknowledgments}

\bibliography{paper_3D}

\end{document}